\documentclass[12pt, letterpaper]{article}

\usepackage{amsmath}
\usepackage{booktabs}
\usepackage{amsthm}
\usepackage{graphicx}
\usepackage[margin=1in]{geometry}
\usepackage{hyperref}
\hypersetup{colorlinks = true, linkcolor = blue, citecolor=blue, urlcolor = blue}
\usepackage{natbib}
\usepackage{enumitem}
\usepackage{setspace}

\usepackage[]{lineno}
\newcommand*\patchAmsMathEnvironmentForLineno[1]{%
 \expandafter\let\csname old#1\expandafter\endcsname\csname #1\endcsname
 \expandafter\let\csname oldend#1\expandafter\endcsname\csname end#1\endcsname
 \renewenvironment{#1}%
 {\linenomath\csname old#1\endcsname}%
 {\csname oldend#1\endcsname\endlinenomath}}%
\newcommand*\patchBothAmsMathEnvironmentsForLineno[1]{%
 \patchAmsMathEnvironmentForLineno{#1}%
 \patchAmsMathEnvironmentForLineno{#1*}}%

\AtBeginDocument{%
 \patchBothAmsMathEnvironmentsForLineno{equation}%
 \patchBothAmsMathEnvironmentsForLineno{align}%
 \patchBothAmsMathEnvironmentsForLineno{flalign}%
 \patchBothAmsMathEnvironmentsForLineno{alignat}%
 \patchBothAmsMathEnvironmentsForLineno{gather}%
 \patchBothAmsMathEnvironmentsForLineno{multline}%
}

\usepackage{xcolor}

\newcommand{\blind}{0}

\begin{document}
%\maketitle

\title{\bf Data Jamboree: A Party of Open-Source Software Solving
  Real-World Data Science Problems}
\if0\blind
{
  \author{Lucy D'Agostino McGowan,
    Shannon Tass,
    Sam Tyner,\\
    HaiYing Wang,
    Jun Yan
  }
} \fi

\maketitle

% \doublespace

\begin{abstract}
The evolving focus in statistics and data science education highlights 
the growing importance of computing. This paper presents the Data 
Jamboree, a live event that combines computational methods with 
traditional statistical techniques to address real-world data science 
problems. Participants, ranging from novices to experienced users, 
followed workshop leaders in using open-source tools like Julia, Python, 
and R to perform tasks such as data cleaning, manipulation, and 
predictive modeling. The Jamboree showcased the educational benefits of 
working with open data, providing participants with practical, hands-on 
experience. We compared the tools in terms of efficiency, flexibility, 
and statistical power, with Julia excelling in performance, Python in 
versatility, and R in statistical analysis and visualization. The paper 
concludes with recommendations for designing similar events to encourage 
collaborative learning and critical thinking in data science.

\end{abstract}

%\doublespace

\section{Introduction}
\label{sec:intro}

% Importance of Computing in Statistics and Data Science Education
The evolving landscape of statistics and data science education increasingly
emphasizes the critical role of computing. This shift, as detailed by
\citet{hardin2021computing} and \citet{nolan2010computing}, reflects a broader
trend towards integrating computational methods with traditional statistical
techniques. These skills are indispensable in modern data analysis, enabling
students to effectively handle and interpret complex datasets. 
\citet{hicks2018guide} further advocate for a comprehensive teaching approach in
data science, one that seamlessly blends computational tools with statistical 
concepts. The inclusion of programming languages like R \citep{R}, Python
\citep{Python}, and Julia \citep{bezanson2017julia} in their
curricula is not just about tool proficiency; it is about fostering a deeper 
understanding of data analysis and statistical inference in a digitally driven 
world.

% Various Forms of Data Competitions and Data Fests
This growing emphasis on computing has led to practical
educational events that give students the opportunity to apply their
computational and statistical skills. Data competitions, such as
hackathons \citep{lara2016hackathons} and DataFests \citep{noll2023insights}, 
represent vital practical components in the education of future data scientists 
and statisticians. While hackathons typically focus on creating a working 
prototype or solution in a short time, often with a strong emphasis on coding 
and immediate problem-solving, DataFests usually involve more in-depth analysis 
of large datasets, focusing on statistical insights and data visualization. 
These events challenge participants to apply their skills in real-world 
scenarios, encouraging not just the application of statistical and 
computational knowledge but also the development of collaborative and 
innovative problem-solving abilities. However, a key limitation of these 
traditional events is that they typically require participants to already have a 
strong background in programming and data science, making it difficult for 
beginners or those with limited experience to fully engage.

% Defining the Data Jamboree
Recognizing the limitations of traditional events, the Data Jamboree was 
developed as a more flexible educational experience, designed to accommodate 
participants with varying levels of expertise. While hackathons and DataFests
also promote inclusivity, mentorship, and hands-on learning, Jamborees emphasize
a workshop-style format that allows participants---especially those with less
technical background---to engage fully from the outset. Led by experienced
mentors, Jamborees foster collaboration across skill levels, ensuring that all
participants can contribute meaningfully regardless of their prior experience in
data science. The Data Jamboree format is flexible and could serve as a
standalone event or as a component within larger events like DataFests,
providing an accessible entry point for participants before engaging in more
complex challenges. It incorporates open-source tools such as 
Julia, Python, and R, allowing participants to explore and compare the strengths 
of each in solving real-world problems. The event’s flexible structure can be
adapted based on
its goals and duration, promoting hands-on learning through active coding
sessions, collaborative problem-solving, 
and guided practice in data science.

% Historical Event Review
The Data Jamboree, organized by the American Statistical Association's (ASA)
Section on Statistical Computing and co-sponsored by the ASA's Section on 
Statistical Graphics, has been held twice, first in November 2022 and again in 
2023. As part of the annual mini-symposium ``Statistical Computing in Action,''
the event gathers data enthusiasts, professionals, and students to engage in 
data cleaning, manipulation, and analysis of open data projects. The Jamboree 
promotes the exchange of knowledge and skills among participants through 
hands-on experiences. The 2023 Jamboree was a one-hour event, the concise format
of which was shaped by the symposium schedule, offering a focused yet impactful
experience that emphasized hands-on learning and collaborative problem-solving.
A unique feature of the event is its focus on using 
multiple open-source tools to collaboratively solve data science problems. 
For example, in 2023, participants used the New York City (NYC) Open Data's 
311 Service Requests (SR) dataset. The event provided opportunities to explore 
and evaluate different programming languages---Julia, R, and Python---enabling
participants to enhance their practical skills while encouraging innovation, 
critical thinking, and teamwork \citep{nolan2010computing, dalzell2023increasing}.

% Advantages of Using Open Data
The use of open data in the Data Jamboree offers both educational and practical 
advantages. Open data initiatives promote transparency and public engagement,
which are essential in data science \citep{beheshti2019datasynapse,
  janssen2012benefits}. By working with open data, participants 
face real-world challenges, using Julia, Python, and R to tackle the same data 
science problems. This hands-on approach not only enhances technical skills but 
also highlights the societal impact of data science. Real-world datasets present 
the messy and complex issues that professionals encounter, offering more 
valuable learning experiences than curated or synthetic datasets often used in 
classrooms. The open data format allows participants to address meaningful issues 
such as public service improvement and urban planning. Additionally, it 
demonstrates the versatility of these tools, showing how each can be used to manipulate, 
analyze, and visualize data differently. The application of these languages in 
real-world scenarios supports the notion that open data can foster diverse, 
data-driven solutions \citep{borgman2012conundrum, cantor2018facets}. Furthermore, 
using open data illustrates how data science contributes to broader societal 
benefits \citep{ridgway2023data}. Ultimately, the Data Jamborees serve as forums 
where participants use open-source software and open data to create valuable 
educational experiences.

% Contributions
The insights gained from the Data Jamboree provide a unique perspective on how
diverse tools like Julia, Python, and R can be applied in educational and
real-world settings. By evaluating the strengths and weaknesses of each tool,
we offer guidance for educators and data science practitioners.
The contributions of this article are three-fold. First, we report on the Data 
Jamboree held at the 2023 Mini-Symposium of the ASA Section on Statistical
Computing, with detailed descriptions of the Jamboree data sets and task,
including the cleaning, manipulation, and analysis of the NYC 311 SR
data. Second, we discuss the use of three popular open-source data science
environments---Julia, Python, and R---within the context of the Jamboree’s
problems, evaluating their strengths and limitations. Third, we provide
recommendations for designing engaging Data Jamboree events that promote data
science outreach and education.

% Organization
The rest of the paper is organized as follows. We first present the dataset and
problems of the Data Jamboree at the 2023 Mini-Symposium on Statistical
Computing in Action, which sets up the arena for the different data science
languages in Section~\ref{sec:Jamboree}. A comparison of Julia, Python, and R
with the major advantages and limitations of each is summarized in
Section~\ref{sec:comp}. Recommendations on how to design a Data Jamboree event
are provided in Section~\ref{sec:design}. A discussion concludes in
Section~\ref{sec:disc}. The code for handling the Jamboree problems using all
three languages is publicly available in GitHub repositories.

\section{The 2023 Data Jamboree}
\label{sec:Jamboree}

The 2023 Data Jamboree attracted over 120 participants, with about half being
students, including 49 student members from the ASA Section on Statistical
Computing. The rest of the attendees were early-career professionals and
experienced data scientists. Participants brought varying levels of familiarity
with R, Python, and Julia, fostering a collaborative environment that encouraged
both peer learning and skill development.

\subsection{Data and Tasks}

The NYC 311 SR dataset represents a comprehensive accumulation of non-emergency
requests made by New York City residents. The dataset includes a wide array of 
urban issues, ranging from noise complaints and pothole reporting to graffiti 
removal and street light issues. Since its inception in 2010, the dataset has 
served as a valuable resource for understanding urban living dynamics and 
informing public service improvements. By providing insight into the spatial and 
temporal patterns of urban issues, the dataset not only reflects the day-to-day 
concerns of New Yorkers but also offers opportunities for analysis that are 
relevant to urban planners and policymakers \citep[e.g.,][]{minkoff2016nyc,
  kontokosta2017equity,  agonafir2022understanding}. The open accessibility of
this dataset makes it an ideal resource for data science projects and
competitions, as it offers both real-world relevance and a wide scope for
exploration.

For the 2023 Data Jamboree, a subset of the NYC 311 SR dataset was selected, 
consisting of requests created between January 15 and 21, 2023. The dataset was 
downloaded in CSV format from the NYC 311 Open Data Portal on February 22, 2023.
This subset contained 54,469 rows and 41 columns, with a total size of 32MB.
Although the subset covers only one week, it still provided participants with 
real-time urban challenges, such as handling incomplete data, testing associations,
and managing large datasets typical of urban environments. This dataset has also
been utilized in various educational settings to address civic questions. For
example, in STAT 3255/5255, Introduction to Data Science at the University of
Connecticut, students explored the NYC 311 data for homework assignments and
midterm projects. Selected students presented their analyses at a workshop
during NYC Open Data Week in March 2023. While independent from the Data
Jamboree, these educational applications highlight the dataset’s versatility and
relevance for real-world data science problems.

The scientific tasks of the Jamboree were divided into three main stages:

\emph{Data Cleaning:} In this stage, participants were required to standardize 
column names across programming languages for ease of comparison, 
identify and correct errors or inconsistencies (e.g., closed dates occurring 
before created dates, invalid values), and handle missing data (e.g., using 
geocoding to fill in missing zip codes). This stage also included a reflection 
task, where participants summarized their suggestions for improvements to the 
data, which were to be communicated to the data curator. These reflections 
ensured that participants not only engaged with data cleaning tasks but also 
considered broader data quality issues.

\emph{Data Manipulation:} Focusing on New York Police Department (NYPD) 
requests, this stage provided specific instructions for participants to create a
new variable,
`duration', representing the time span between the Created Date and the Closed
Date. Participants were then requested to visualize the distribution of the `duration' 
variable across weekdays, weekends, and boroughs, and to test for similarities 
in these distributions. Additionally, they were instructed to merge zip code-level 
data, such as population density, home values, and household income from the 
US Census, with the NYPD requests data for further analysis. This added dimension 
allowed for a more nuanced exploration of the socioeconomic factors influencing 
the handeling of NYPD service requests. The structured approach ensured
participants could successfully complete foundational tasks, setting the stage
for more open-ended analysis in subsequent stages.

\emph{Data Analysis:} The final stage required participants to define a binary 
variable, `over3h', indicating whether the time to close a NYPD SR exceeded 
three hours. They were tasked with using logistic regression models to predict
this variable, incorporating 311 SR data and zip code-level covariates. 
Participants were also encouraged to apply alternative models, such as random 
forests or neural networks, to explore different analytical approaches. This 
diversity of modeling techniques allowed participants to compare the performance 
and interpretability of various methods, enhancing their understanding of 
predictive modeling in real-world data contexts.

Additional details on the Jamboree's tasks are available at 
\url{https://asa-ssc.github.io/minisymp2023/jamboree/}.

\subsection{Workshops}

The 2023 Data Jamboree featured three of the most widely used open-source 
programming languages in data science: Julia, Python, and R. Each language was 
selected for its strengths in different aspects of data science. Python, 
known for its versatility and broad array of libraries, is widely used in 
machine learning and general-purpose data manipulation \citep{vanderplas2018python}. 
R, on the other hand, is highly regarded for its statistical analysis and 
visualization capabilities \citep{wickham2017r}, making it a favorite among 
statisticians and data scientists alike. Julia, with its growing user base, 
has gained popularity for its high performance in handling large datasets 
\citep{bezanson2017julia}, positioning itself as a strong competitor in 
scientific computing and data-heavy tasks. These languages have been compared 
in various applications for their reproducibility and performance 
\citep[e.g.,][]{stanish2023reproducibility}.

The workshops were featured events during the Data Jamboree, and while
attendance was voluntary, they were central to the event’s
structure. Participants were advised to pre-install Julia, Python, and R ahead
of the event to ensure smooth participation. However, installation of add-on
packages was managed during the workshops, as all three languages provide
convenient package management systems that allow for quick and easy
installation.
Each language was allocated 15 minutes to address the same tasks using the NYC
311 data. During these sessions, the leader presented their solutions on screen
while participants followed along on their own computers using code shared via
GitHub repositories. This follow-along structure ensured that participants could
engage directly with the material during the workshop. The flexible nature of
this format, as described in Section~\ref{sec:design}, allows organizers to
adjust the event length and level of detail based on their time budget and
participant engagement. After the demonstrations, a question-and-answer session
allowed participants to explore the tools further.

The Julia, Python, 
and R workshops were led by HaiYing Wang, Shannon Taas, and Lucy McGowan, 
respectively. The Jamboree was moderated by Sam Tyner, who had led the R workshop 
during the 2022 Data Jamboree. Video recordings of the workshops are publicly 
available on the YouTube channel of the ASA Section on Statistical Computing 
and ASA Section on Statistical Graphics 
(\url{https://www.youtube.com/@statgraphics}). The codes used by the workshop 
leaders are also publicly available in GitHub repositories.

\section{Comparison of Julia, Python, and R}
\label{sec:comp}

The Data Jamboree provided an ideal platform to compare three popular open-source 
languages---Julia, Python, and R---within the context of real-world data science tasks. 
These tools were evaluated based on their performance in handling the NYC 311 SR 
dataset during the Jamboree, specifically focusing on data cleaning, manipulation, 
and analysis. The comparison takes into account each tool’s efficiency in managing 
large, complex data, their flexibility in communicating with other tools, and 
their power in performing statistical analysis and visualization. Furthermore, 
this evaluation considers the educational impact of these tools, particularly 
their accessibility for learners with different skill levels and their ability to 
support hands-on learning experiences in data science. By framing this comparison 
around the practical tasks completed by participants during the Jamboree, we 
highlight the strengths and limitations of Julia, Python, and R in both professional 
and educational settings.

\subsection{Julia}

The Julia solution to the Jamboree problems by HaiYing Wang is publicly
available at
\url{https://github.com/Ossifragus/DataJamboree}.
Julia’s just-in-time (JIT) compilation offers advantages in
processing large, complex datasets, making it highly efficient. For example,
during the Data Jamboree, Julia efficiently managed the extensive NYC 311 SR data,
performing aggregations and joins to analyze the distribution and frequency of
various SR across different boroughs. The ability to compile code
to machine language allows Julia to execute these tasks at speeds comparable to
traditional high-performance languages like C and Fortran.

One of Julia’s key strengths is its flexibility in integrating with other tools, 
allowing participants to access the powerful capabilities of Python and R. 
Through the \textsf{PythonCall} package \citep{PythonCall.jl},
Julia can call Python’s machine learning 
libraries, such as TensorFlow and Scikit-learn, which was especially useful 
during the Data Jamboree when we analyzed the NYC 311 SR data. 
This enabled them to apply advanced machine learning models for predicting trends 
in SR, while leveraging Julia’s computational efficiency for 
handling large datasets. Additionally, Julia’s \textsf{RCall} package {RCall.jl}
provides
access to R’s specialized statistical methods and visualization tools, such as 
\textsf{ggplot2} \citep{ggplot2},
further enhancing the analysis of temporal and spatial patterns in the 
data. By integrating Python’s machine learning capabilities and R’s statistical 
tools, Julia offered participants a versatile environment to perform complex 
data manipulation, analysis, and visualization tasks on the NYC 311 SR dataset.

Julia also provides a first-class reproducibility system that stands out
compared to R and Python, thanks to its native \textsf{Pkg} system \citep{Pkg.jl}. Unlike
external tools in other languages, Julia’s package management and environment
isolation are deeply embedded within the language itself, with a strong focus on
reproducibility from the ground up. The \textsf{Pkg} system ensures every
project operates in its own isolated environment, with specific package versions
that are fully insulated from other projects. This greatly minimizes dependency
conflicts and makes it easy to recreate identical environments across different
systems, enhancing both reliability and collaboration.

In terms of data manipulation and computational tasks, Julia offers powerful
capabilities through packages like \textsf{DataFrames.jl}
\citep{BouchetValat2023}. These libraries provide
comprehensive functionalities for handling large datasets and performing complex
transformations essential for exploratory data analysis. During the Data
Jamboree, we utilized \textsf{DataFrames.jl} to efficiently filter, sort, and
aggregate the NYC 311 SR data. For instance, we analyzed the temporal trends of
different types of SR and identified peak times for noise
complaints, enabling more precise analysis and reporting.

However, Julia's ecosystem is not as mature as that of Python or R. The
availability of specialized libraries and community support for niche areas can
be limited. During the Data Jamboree, we faced challenges finding
well-documented libraries for specific tasks, such as geospatial analysis, which
are readily available in Python or R. This limitation can slow down the
workflow, requiring users to develop custom solutions or bridge Julia with other
languages.

Additionally, Julia has a steeper learning curve compared to Python. While
praised for its performance, new users had to spend additional time learning its
syntax and understanding its ecosystem. This initial learning investment can be
a barrier in time-constrained environments like hackathons, where quick
onboarding and rapid development are crucial. Novices could struggle with
the syntax differences when transitioning from more familiar environments,
impacting their productivity initially.

Integrating Julia into existing data science workflows that predominantly use
Python or R can also present challenges. Despite tools like \textsf{PyCall}
\citep{PyCall.jl} and \textsf{RCall} \citep{RCall.jl} facilitating
integration, differences in syntax, data handling, and library functionalities
require additional adjustments. During the Data Jamboree, we needed to
create wrappers and compatibility layers to ensure smooth interoperability
between Julia and other languages. For instance, integrating visualization
outputs from Julia with a dashboard created in Python’s \textsf{Dash}
\citep{Dash.jl} required careful
handling of data formats and function calls, introducing complexity and
potential points of failure in the workflow.

In conclusion, Julia’s performance during the Data Jamboree highlighted its
potential and limitations. Its ability to efficiently handle large datasets,
flexibility in integrating with other tools, and strong capabilities for
statistical analysis and visualization make it a powerful tool for data
science. However, the relatively immature ecosystem, steeper learning curve, and
integration challenges with established workflows are notable drawbacks. As
Julia continues to evolve, it is poised to become a more integral part of the
data science toolkit, particularly for tasks demanding high computational
performance.

\subsection{Python}

The Python solution to the Jamboree problems by Shannon Tass is publicly
available at \url{https://github.com/esnt/Jamboree}.
Python’s extensive library ecosystem and user-friendly syntax made it a
popular choice among participants for various data science tasks. The Jamboree
showcased Python's strengths in data manipulation, integration with other tools,
and data visualization, while also revealing some of its limitations.

Python’s versatility is one of its key strengths, supported by a rich ecosystem
of libraries designed for data manipulation, analysis, and visualization. During
the Data Jamboree, we leveraged \textsf{Pandas} \citep{mckinney2010data}, a powerful data manipulation
library, to handle the NYC 311 SR data efficiently. Tasks such as data cleaning,
filtering, and aggregation were streamlined using \textsf{Pandas}’ intuitive syntax. For
example, participants used \textsf{Pandas} to filter SR by borough and
service type, enabling a detailed analysis of complaint patterns across
different areas of New York City.

In addition to data manipulation, Python excels in integrating with various
tools and environments, enhancing its utility in diverse workflows. The ability
to call R functions using the \textsf{rpy2} package \citep{rpy2}, for instance, allowed participants
to utilize R’s statistical packages within a Python environment. This was
particularly useful for performing advanced statistical tests on the NYC 311 SR
data, where Python was used to handle the data preparation and visualization,
and R was used to
conduct the statistical analysis. Such integration capabilities underscore
Python’s flexibility in combining strengths from different programming
languages.

Python’s data visualization capabilities are another advantage, with
packages like \textsf{Matplotlib} \citep{hunter2007matplotlib} and
\textsf{Seaborn} \citep{waskom2021seaborn} providing tools for creating
detailed and informative visualizations. During the Data Jamboree, we
used these libraries to visualize trends and distributions in the NYC 311 SR
data. For instance, \textsf{Seaborn} was used to create heatmaps showing the density of
SR in different neighborhoods, while \textsf{Matplotlib} helped in plotting
time series to analyze trends in complaint types over time. These visualizations
were crucial for identifying patterns and drawing actionable insights from the
data.

However, Python also has its limitations. One of the challenges one
faces is performance issues when handling extremely large datasets. While
libraries like \textsf{Pandas} for data frames are powerful, they are not
optimized for very large-scale
data processing, leading to slower performance compared to languages like
Julia. For example, aggregating millions of records in the NYC 311 SR dataset
sometimes resulted in significant processing times, highlighting the need for
performance optimization techniques or the use of more specialized tools like
the package \textsf{Dask} \citep{dask}.

Another limitation is Python’s less mature support for certain statistical and
mathematical operations compared to R. While Python’s libraries are
comprehensive, they may not offer the same depth of specialized statistical
functions available in R. During the Jamboree, we occasionally needed
to switch to R for specific statistical analyses that were more straightforward
to perform using R’s specialized packages. This dual-language workflow, although
effective, added complexity to the process.

Python’s learning curve, while generally considered accessible, can still pose
challenges for complete beginners. Participants new to programming or data
science had to invest time in learning Python’s syntax and understanding its
diverse library ecosystem. Despite its readability and user-friendly nature, the
initial learning phase can be daunting, especially in a high-pressure
environment like a hackathon.

In conclusion, Python’s performance during the Data Jamboree highlighted its
versatility and extensive library support for data manipulation, integration, and
visualization. Its ability to combine strengths from different programming
languages through seamless integration makes it a powerful tool for data
science. However, performance limitations with very large datasets, less mature
statistical support compared to R, and the initial learning curve are notable
drawbacks. As Python continues to evolve, it remains a fundamental tool in the
data science toolkit, particularly for tasks requiring extensive data
manipulation and visualization.

\subsection{R}

The R solution to the Jamboree problems by Lucy D’Agostino McGowan is publicly
available at
\url{https://github.com/LucyMcGowan/2023-data-Jamboree}.
The Jamboree highlighted R's powerful
data manipulation tools, integration with other languages, and advanced
visualization capabilities, while also revealing some of its limitations.

R's strength in data manipulation was evident as we used the
\textsf{dplyr} \citep{dplyr} and \textsf{tidyr} \citep{tydyr}
packages to clean and transform the NYC 311 SR data. These packages allowed
for streamlined data wrangling, making it easy to filter, sort, and aggregate
the data. For example, we used \textsf{dplyr} to group SR by type
and borough, summarizing the frequency of different complaint types across New
York City. The \textsf{tidyverse} suite \citep{wichham2019welcome}, which
includes \textsf{dplyr} and \textsf{tidyr}, provides a
cohesive set of tools with a shared syntax and grammar that simplify complex
data manipulation tasks, making R a powerful language for data scientists.

Another advantage of R is its advanced statistical analysis
capabilities. R has a rich ecosystem of packages for performing a wide range of
statistical tests and models. During the Data Jamboree, we used standard
packages like \textsf{stats} and \textsf{MASS} \citep{venables2002modern} to
conduct hypothesis testing and fit regression
models. For instance, we analyzed the correlation between complaint types and
response times, using R's statistical tools to derive meaningful insights from
the data. R's extensive collection of widely used specialized packages, such as
\textsf{survival} \citep{therneau2000modeling} for survival analysis and
\textsf{lme4} \citep{bates2015fitting} for mixed-effects models, enables users to perform
advanced statistical analyses that are crucial in many data science projects.

R's visualization capabilities are another area where it excels. The \textsf{ggplot2}
package \citep{ggplot2}, also part of the \textsf{tidyverse}, is renowned for its ability to create
high-quality, customizable visualizations. We used \textsf{ggplot2} to create
detailed plots that showcased trends and patterns in the NYC 311 SR data. For
example, we generated heatmaps to visualize the density of SR in
different neighborhoods and time series plots to analyze the trends in
complaints over time. These visualizations helped us identify
patterns and communicate the findings effectively.

However, R also has its limitations. One challenge is performance when handling
very large datasets. While R is highly efficient for many data manipulation
tasks, it can struggle with memory management and processing speed for datasets
approaching one-tenth of computer memory.
For the 2023 Data Jamboree tasks, one could experience slow
performance if working with the full NYC 311 SR dataset, necessitating the use of
data sampling or external data storage solutions to manage memory usage
effectively.

Another limitation is R's integration with other programming environments. While
R can call functions from other languages using interfaces like
\textsf{reticulate} \citep{rticulate} for Python and \textsf{Rcpp}
\citep{eddelbuettel2013seamless} for C++ , these integrations can sometimes be
cumbersome. One could face
challenges when integrating R with external tools or other programming
languages, adding complexity to their workflows.

Lastly, R can have a steep learning curve for users who are new to programming
or data science. While R's syntax is highly expressive and well-suited for
statistical analysis, it can be less intuitive for beginners. The tidyverse
suite of packages seeks to bridge this gap, but many users have to supplement
their analyses with other packages with differing syntax. Those who are new to R
may have to invest additional time learning the syntax and understanding its
extensive package ecosystem, which could be a barrier in fast-paced hackathon
environments.

In conclusion, R's performance during the Data Jamboree highlighted its
strengths in data manipulation, statistical analysis, and visualization. Its
cohesive set of tools and specialized packages make it a powerful language for
data scientists. However, challenges with performance on large datasets,
integration with other environments, and a steeper learning curve are potential
drawbacks. As R continues to evolve, it remains an essential tool in the data
science toolkit, particularly for projects requiring advanced statistical
analysis and high-quality visualizations.

\section{Design Your Own Jamboree}
\label{sec:design}

Designing a successful Data Jamboree requires careful planning and several key 
considerations to ensure an engaging and productive experience for participants. 
Below are essential guidelines to help structure the event effectively.

\paragraph{Defining the Aims}

The first step is to clearly define the objectives of the event. Identify your 
target participants—students, data professionals, or a combination of both—and 
determine what they should gain from the event. Will the focus be on general 
education, skill-building, or performance comparisons across different tools? 
Establish whether the Jamboree will emphasize statistical analysis, predictive 
modeling, or data visualization. These decisions will guide the event's structure 
and content. The Data Jamboree format is highly adaptable and can
function as a standalone event or as a featured component within larger data
science gatherings, such as DataFests. Integrating a Jamboree into a DataFest
could serve as a preparatory workshop, providing participants with hands-on
experience before engaging in more complex data challenges.

\paragraph{Selecting the Data}

Choosing the right datasets is pivotal to the success of the Jamboree. 
Leveraging open data initiatives provides access to diverse and meaningful 
datasets. Examples include NYC Open Data, which offers a wide range of public 
datasets from various city agencies, data.gov for nationwide datasets across 
different sectors, NOAA for comprehensive climate data, and the U.S. Census 
Bureau for demographic data. A well-chosen dataset that is rich and complex can 
support a variety of data science tasks, such as data cleaning, curation, and 
statistical analysis. Utilizing datasets from multiple sources introduces 
additional challenges, particularly in terms of data integration and 
manipulation, which mirrors real-world scenarios. This complexity not only 
enhances participants' technical skills but also provides a more comprehensive 
understanding of data science workflows.

\paragraph{Designing the Jamboree Questions}

The questions posed to participants should cover a range of data science 
tasks, such as data cleaning, exploratory analysis, statistical testing, 
predictive modeling, and visualization. This ensures that participants can 
apply a wide range of skills during the event. For example, tasks may involve 
identifying and rectifying data errors, exploring trends in the data, conducting 
statistical tests, or building machine learning models to predict outcomes. 
These questions should be aligned with the event's overall aims and the complexity 
of the chosen data.

\paragraph{Determining the Event Length}

The event length should align with the complexity of the tasks and the
organizers' time budget. While our Jamboree was completed in one hour, higher
levels of participant engagement, dedicated software setup time,
extended question and answer sessions, or more in-depth tasks---such as machine 
learning or predictive modeling---may necessitate a half-day or even a full-day 
event. The leader presentations, and participant activities are
integrated into a single session, with participants following along as the leaders
present solutions. This format provides flexibility, allowing organizers to
adjust both the duration and depth of the event based on available time and
audience needs.
For longer events, logistical considerations like meal breaks and lodging 
become important, especially for participants traveling from afar. Ensuring access 
to essential facilities, including a reliable internet connection and adequate power outlets, 
is critical for maintaining productivity and focus. In virtual events, certain
logistical challenges are reduced, but collaboration and communication can be
more difficult, potentially affecting engagement. The duration should still reflect the
difficulty of the tasks to ensure a balance between engagement and content 
coverage.

\paragraph{Finding Workshop Leaders}

Experienced workshop leaders play a pivotal role in ensuring the success of 
the Jamboree. Leaders should have expertise in both data science and the 
specific tools being used, such as R, Python, or Julia. They should guide 
participants through the tasks, offering insights and fostering collaboration. 
We recommend having at least one leader per programming language or key task,
with a student-to-instructor ratio that ensures adequate support for all
participants, including teaching assistant if resources allow. Leaders may 
come from academic institutions, industry, or data science communities.

\paragraph{Special Considerations for Minors}

If minors are involved, additional steps must be taken to ensure their safety 
and compliance with legal requirements. This includes obtaining parental 
consent, designing age-appropriate challenges, and providing adequate supervision. 
Any legal or institutional requirements must be adhered to, such as ensuring 
leaders have undergone background checks or assigning designated guardians 
during the event. Addressing these considerations helps create a safe and 
welcoming environment for younger participants.

\paragraph{Post-Event Analysis and Feedback}

Post-event analysis is essential for refining future Jamborees. We recommend 
distributing surveys to gather feedback from participants about their
experience. To maximize response rates, feedback forms should be administered
immediately after the event.
Additionally, holding a debrief session for leaders allows for the identification 
of what worked well and where improvements can be made. Documenting these insights 
will help establish a foundation of knowledge that can be shared with future 
organizers. Over time, this feedback will contribute to an evolving institutional 
knowledge that enhances the quality of the event and ensures its continued success.

\bigskip

These recommendations provide a comprehensive framework for organizing a successful 
Data Jamboree. By focusing on clear objectives, appropriate datasets, well-designed 
questions, event logistics, and thoughtful post-event analysis, organizers can 
create an engaging and educational experience for participants. Ensuring a balance 
between structure and flexibility fosters both innovation and collaboration.

\section{Discussion}
\label{sec:disc}

Comparing Julia, Python, and R, we find each language has unique strengths and
limitations that impact their effectiveness in a data science educational
setting. Julia stands out for its high performance, particularly with large
datasets, due to its just-in-time (JIT) compilation. Additionally, Julia can
seamlessly call Python libraries using the \textsf{PyCall} package
\citep{PyCall.jl} and R libraries using \textsf{RCall} \citep{RCall.jl},
enhancing its versatility. Python is celebrated for its extensive library
support and accessibility, making it powerful for a wide range of applications,
particularly in deep learning with libraries like \textsf{TensorFlow}
\citep{tensorflow2015}, \textsf{Keras} \citep{chollet2015keras}, and
\textsf{PyTorch} \citep{paszke2019pytorch}.
R excels in statistical analysis and visualization, with a comprehensive
ecosystem of packages tailored for these purposes. Unified data formats, such as
Arrow \citep{bates2024csv}, can further streamline cross-platform data
manipulation and enhance
computational efficiency across Julia, Python, and R.

Notably, we did not include commercial software such as SAS, Matlab, or
Mathematica in our comparison. These tools, while powerful, are less accessible
due to licensing costs and do not align with the open-source philosophy that
fosters broad educational engagement. Furthermore, we do not recommend
tools like MS Excel for data science. While Excel can serve as an accessible
entry point for basic data manipulation and visualization, it may present
challenges for advanced data analysis due to its limitations in accuracy and
suscesptibility to user errors \citep{mccullough2008special,
  rajalingham2016limitations}. 
Additionally, reliance on Excel may encourage poor practices such as
manual data manipulation, lack of reproducibility, and overuse of hard-coded
values, which hinder the scalability and transparency of analyses. Time invested
in learning Excel could be more effectively used to master specialized tools
like Julia, Python, and R, which offer greater capabilities for data science.

The Data Jamboree format is broadly applicable to events like DataFests and 
other educational activities, emphasizing the value of tackling real-world 
problems with open data. Unlike hackathons, the workshop style of the Data 
Jamboree encourages participation from all skill levels, not just those with 
coding experience. Using real-world datasets, such as those
provided by Open Data initiatives, allows participants to tackle practical
challenges, enhancing their problem-solving skills and understanding of data
science applications. This approach not only fosters engagement and excitement
but also prepares students for professional scenarios where they must analyze,
interpret, and make decisions based on real-world data. Incorporating these
elements into educational events ensures a comprehensive learning experience
that bridges the gap between theoretical knowledge and practical application.

To further enhance the educational impact of future Data Jamborees,
incorporating systematic performance evaluation and participant feedback will be
valuable. While the 2023 Data Jamboree did not focus on detailed
profiling of computational performance across programming languages, future
events could incorporate systematic benchmarking as part of the analysis. This
would provide participants with deeper insights into the efficiency and
scalability of different tools, particularly when working with larger datasets
or more complex tasks. Future Data Jamborees could also benefit from
systematically collecting data on participant task completion rates and
experience levels. This information would offer valuable insights into how
different audiences engage with the material and could guide adjustments in
pacing, content depth, and workshop structure to better meet participant needs.

\bibliographystyle{apalike}
\bibliography{citations}

\end{document}